\documentclass[12pt]{article}
\usepackage{amsfonts}
\title{Construction of the effective action in
nonanticommutative supersymmetric field theories}

\author{O.D. Azorkina\footnote{azorkina@tspu.edu.ru}\, $^1$,
A.T. Banin\footnote{atb@math.nsc.ru}\, $^2$, I.L.
Buchbinder\footnote{joseph@tspu.edu.ru, J.Buchbinder@damtp.cam.ac.uk}\, $^3 \,
\footnote{On leave of absence from Department of Theoretical Physics,
Tomsk State Pedagogical University, Tomsk 634041, Russia}$, N.G.
Pletnev\footnote{pletnev@math.nsc.ru}\, $^2$}

\date{\it
$^1$ Department of Theoretical Physics\\ Tomsk State Pedagogical
University\\ Tomsk 634041, Russia\\$^2$ Department of Theoretical
Physics\\ Institute of Mathematics, Novosibirsk, \\ 630090,
Russia\\$^3$ Department of Applied Mathematics and Theoretical
Physics\\ University of Cambridge, Centre for Mathematical
Sciences\\ Wilberforce Road, Cambridge, CB3 0WA, UK }

\begin{document}
\maketitle

\vspace{-12cm}
\begin{flushright}
DAMTP-2005-92\\
hep-th/0509193
\end{flushright}
\vspace{10.0cm}

\begin{abstract}
We develop a general gauge invariant construction of the
one-loop effective action for supersymmetric gauge field theories
formulated in ${\cal N}=1/2$ superspace. Using manifestly
covariant techniques (the background superfield method and
proper-time representations) adopted to the ${\cal N}=1/2$
superspace we show how to define unambiguously the effective
action of a matter multiplet (in fundamental and adjoint
representations) and the vector multiplet coupled to a background
${\cal N}=1/2$ gauge superfield. As an application of this
construction we exactly calculate the low-energy one-loop
effective action of matter multiplet and $SU(2)$ SYM theory on the
Abelian background.

\end{abstract}
\thispagestyle{empty}
\newcommand{\be}{\begin{equation}}
\newcommand{\ee}{\end{equation}}
\newcommand{\bea}{\begin{eqnarray}}
\newcommand{\eea}{\end{eqnarray}}

\section{Introduction}
Recently it was shown \cite{kl} that the low-energy limit of
superstring theory in the self-dual graviphoton background field
$F^{\alpha\beta}$ leads to a four-dimensional supersymmetric field
theory formulated in the deformed ${\cal N}=1$ superspace with
fermionic coordinated satisfying the relation
\begin{equation}
\label{hat}
\{\hat{\theta}^\alpha, \hat{\theta}^\beta\}=2 \alpha^{'\,
2}F^{\alpha\beta}=2{\cal C}^{\alpha\beta}.
\end{equation}
The anti-commutation relations of the remaining ${\cal N}=1$
chiral superspace coordinates $y^m, \bar{\theta}^{\dot\alpha}$ are
not modified. The field theories defined on such superspace can be
formulated via ordinary superfield actions where the superfields
product is defined via the star product \cite{S}
\begin{equation}
\label{star} f(x,\theta,\bar{\theta}) \star g
(x,\theta,\bar{\theta}) = f {\rm e}^{\overleftarrow{Q_\alpha}
{\displaystyle\cal C}^{\alpha\beta} \overrightarrow{Q_\beta}}g.
\end{equation}
Here $Q_\alpha =i\frac{\partial}{\partial \theta^\alpha} +
\frac{1}{2}\bar{\theta}^{\dot\alpha}\partial_{\alpha\dot\alpha}$
is the supersymmetry generator. Since the star-product contains
explicitly the supersymmetry generator $Q_{\alpha}$ a half of
supersymmetries is broken down. Therefore this superspace is
called ${\cal N}=1/2$ superspace and the corresponding field
theories are called ${\cal N}=1/2$ or nonanticommutative (NAC)
supersymmetric theories. We emphasize that such a superspace
deformation is possible only in the Euclidean space\footnote {In
the case of extended supersymmetric theories, the various
superspace deformations can be constructed in the sector of
fermionic coordinates \cite{Iv}.}.

Renormalization of various ${\cal N}=1/2$ supersymmetric theories
was discussed in Refs.\cite{7}, \cite{8}.  One-loop calculations
of counterterms in supergauge models were carried out in component
formalism \cite{9} and in superspace \cite{10}. It has been found
\cite{9}, that at one loop, in the standard class of gauges, the
non-gauge-invariant divergent terms are generated. However, there
exists a non-linear redefinition of the gaugino fields which
remove such terms and restores the gauge invariance. Moreover a
modified version of the original pure ${\cal N}=1/2$ Lagrangian
has been proposed in \cite{9}. It has a form preserved under
renormalization and the parameter of
nonanticommutativity ${\cal C}^{\alpha\beta}$ is unrenormalized
(at least at one loop). The one loop divergences for NAC $U(N)$
gauge theories with the matter in the adjoint representation have been
studied in Ref. \cite{10} using superfield background field
method. It was found that the divergent non-gauge -invariant
contributions are generated, however it was proved that sum of all
one-loop divergences is gauge invariant without any redefinitions.
We point out that practically all results in ${\cal N}=1/2$ gauge
theories concerned only structure of divergences, the finite part
of effective action has not been investigated.

The gauge invariance of the above results means that at the
quantum level supergauge invariance is consistent with NAC
geometry and hence there exists a procedure to construct the
effective action in a form which preserves the manifestly gauge
invariance and ${\cal N}=1/2$ supersymmetry. The aim of this paper
is to formulate the procedure and study some of its
applications. We consider ${\cal N}=1/2$ SYM theory coupled to
a matter in fundamental and adjoint representations and use the
superfield background field method (see e.g. \cite{11}) together
with superfield heat kernel method \cite{12}. For evaluation of
one-loop effective action we also use the special techniques
developed in Refs. \cite{13} adapting it to ${\cal N}=1/2$
superspace.

The paper is organize as follows. In the Section 2 we formulate
the basic properties of nonanticommutative star-product and
introduce the operators $T^{*}_{c}$ and $T^{*}_{s}$
which are very important for considering the NAC matter in
the adjoint representation and pure NAC SYM theory. Section 3 is
devoted to formulation of the models in superspace and superfield
background field method. In the Section 4 we describe a
calculation of the one-loop effective action for the matter in the fundamental
representation coupled to an external SYM field and find the one-loop
effective action on a covariantly constant on-shell background.
Section 5 is devoted to discussion of the heat kernel techniques and the
one-loop effective action for the NAC SYM theory. The obtained results are
formulated in the Summary. We do not discuss the details of
calculations which are analogous to ones in conventional
superfield theories (see e.g \cite{11}, \cite{12}, \cite{13},
\cite{16}, \cite{18}) and pay attention only on the aspects
essentially associated with $\star$-operation.

\section{The properties of $\star$-product}

The field theories on the ${\cal N}=1/2$ superspace can be
conveniently formulated using a notion of symbols of the
operators. For any operator function $\hat{f}$ depending on
variables $\hat{\theta^\alpha}$ satisfying the relations
(\ref{hat}) one defines the corresponding Weyl symbol $f(\theta)$
by the rule $\hat{f}=\int d^2\pi {\rm e}^{\pi\hat{\theta}}
\tilde{f}(\pi) $ where $\tilde{f}(\pi)$ is the Fourier transform
of the symbol $f(\theta)$: $\tilde{f}(\pi)=-\int d^2\theta {\rm
e}^{-\pi\theta}f(\theta)$ (see some details in \cite{alv}). The
delta function of the anticommuting variables $\theta$ is
presented as $\delta(\theta-\theta')= \int d^2 \pi {\rm
e}^{\pi(\theta-\theta')} .$ The set of operator functions forms a
graded algebra. The in product of two operators $\hat{f}, \hat{g}$
is associated with the star-product of the corresponding Weyl
symbols $\hat{f} \cdot \hat{g} =\int d^2\pi {\rm
e}^{\pi\hat{\theta}}(\widetilde{f\star g})(\pi),$ where the
star-product is defined by
\begin{equation}\label{st}
f(\theta)\star
g(\theta)=f(\theta)e^{-\overleftarrow{\frac{\partial}{\partial\theta^\alpha}}\,{\displaystyle\cal
C}^{\alpha\beta}\,\overrightarrow{\frac{\partial}{\partial\theta^\beta}}}g(\theta)=f\cdot
g -(-1)^{|f|}\frac{\partial f}{\partial \theta^\alpha}{\cal
C}^{\alpha\beta}\frac{\partial g}{\partial \theta^\beta}
-\frac{1}{2}\frac{\partial^2 f}{\partial \theta^2}{\cal C}^2
\frac{\partial^2 g}{\partial \theta^2}.
\end{equation}
A star-product of exponential factors ${\rm e}^{\theta\pi}\star
{\rm e}^{\theta\psi} = {\rm e}^{\theta\pi +\theta\psi -\pi_\alpha
{\cal C}^{\alpha\beta}\psi_\beta}$ allows to write a star-product
for two Weyl symbols in the form $f \star g=\int d^2 \pi\, {\rm
e}^{\pi \theta} f(\theta^\alpha +{\cal
C}^{\alpha\beta}\pi_\beta)\tilde{g}(\pi) ~,$ as well as the
star-products of symbols and delta-function in the form
\begin{equation}
f_1 \star \ldots\star  f_{n}\star \delta(\theta-\theta')\star g_1
\star \ldots \star g_m=
\end{equation}
$$
\int d^2 \pi\, {\rm e}^{(\theta-\theta') \pi} f_1(\theta+{\cal
C}\pi)\star \ldots \star f_n(\theta+{\cal C}\pi) \star
g_1(\theta-{\cal C}\pi)\star \ldots \star g_m(\theta-{\cal
C}\pi)~.
$$
One can note that the change of variables $\theta \rightarrow
\theta \pm {\cal C}\pi$ in the chain of the left (right)
star-products of symbols and delta-functions does not affect on
the exponent argument because of the property $\pi{\cal C}\pi=0$,
however it simplifies arguments of $f_{i}$, for example:
\begin{equation}\label{main} f_1 \star
\ldots\star f_{n-1}\star f_{n}\star \delta(\theta-\theta') = \int
d^2 \pi\, {\rm e}^{(\theta -\theta')\pi} f_1(\theta)\star \ldots
\star f_{n-1}(\theta)\star f_n(\theta)~.
\end{equation}
These are the
basic properties of the star-products which allows to adopt the rules
of operation with superfields  in the conventional superspace theory for the NAC
superspace\footnote{See some
analogous rules in noncommutative non-supersymmetric field theory in \cite{19}}.

Using the star-product operation (\ref{st}) we define the
star-operators $T^{\star}_{c}$ and $T^{\star}_{s}$  by the rule
\begin{equation}\label{t}
T^\star_c=\frac{1}{2}(\star +\star^{(-1)}
)=\cosh(\overleftarrow{\frac{\partial}{\partial\theta^\alpha}}{\cal
C}^{\alpha\beta}\overrightarrow{\frac{\partial}{\partial\theta^\beta}}),\quad
T^\star_s=\frac{1}{2}(\star -\star^{(-1)}
)=\sinh(-\overleftarrow{\frac{\partial}{\partial\theta^\alpha}}{\cal
C}^{\alpha\beta}\overrightarrow{\frac{\partial}{\partial\theta^\beta}})
\end{equation}
where $\star^{(-1)}$ means (\ref{st}) with replacement ${\cal
C}\rightarrow -{\cal C}$. There is the Leibniz rule for
$T^\star_{c,s}$ products $\nabla_A(fT^\star g)=(\nabla_A f)T^\star
g +fT^\star \nabla_A g$ and the Jacoby identity $fT^\star g
T^\star h+hT^\star f T^\star g +gT^\star h T^\star f=0$ for
integrand. It is easily to understand that any chain of
$T^\star_s$-products is the total derivative
\begin{equation}\label{sinh} f T^\star_s g
=(-1)^{|f|}\partial_\alpha(f {\cal C}^{\alpha\beta} \partial_\beta
g).\end{equation} Further we will see that the operators
$T^{*}_{c}$ and $T^{*}_{s}$ are very important for evaluation of
effective action for fields in adjoint representation.

\section{SYM theory coupled to the chiral matter in ${\cal N}=1/2$ superspace
and the background field method}

The ${\cal N}=1/2$ SYM theory in four dimensional superspace with explicitly
broken supersymmetry in the antichiral sector can be defined
on a ${\cal N}=1$ superspace \cite{S} as straightforward
generalization of the standard construction (see e.g. \cite{11},
\cite{12}) by introducing the NAC (but associative) star-product
(\ref{star}). The gauge transformations of (anti)chiral
superfields in fundamental representation are given in terms of
two independent chiral and antichiral parameter superfields
$\Lambda, \bar{\Lambda}$ as follows
$$
\Phi'={\rm e}^{i\Lambda}_\star \star \Phi, \quad \bar{\Phi}'=
\bar{\Phi} \star {\rm e}^{-i\bar{\Lambda}}_\star.
$$
Gauge fields and field strengths
together with their superpartners can be organized into
superfields, which are expressed in terms of a scalar superfield
potential $V$
in the adjoint representation of the gauge group.
The gauge transformations of $V$ look like
$$
{\rm e}^{V'}_\star= {\rm e}^{i\bar{\Lambda}}_\star \star {\rm
e}^V_\star \star {\rm e}^{-i\Lambda}_\star.
$$
Studying of component structure of supergauge theories is simplified
with the help of Wess-Zumino (WZ) gauge. It was shown \cite{S}
that the commutative gauge transformation does not preserve the WZ
gauge because of the properties of $\star$-product and one needs
to perform an additional ${\cal C}$-dependent gauge transformation
in order to recover the WZ gauge. Therefore the supersymmetry
transformations of the component fields receive a deformation
stipulated by the parameter of nonanticommutativity ${\cal
C}^{\alpha\beta}$.

In the gauge chiral representation the constraints for the
superspace covariant derivatives are solved by
\begin{equation}\label{nabla}
\nabla_A=D_A-i\Gamma_A =({\rm e}^{-V}_\star \star D_\alpha {\rm
e}^V_\star, \bar{D}_{\dot\alpha}, -i\{\nabla_\alpha, \star
\bar{\nabla}_{\dot\alpha}\})
\end{equation}
and the corresponding
superfield strengths are given by the deformed algebra of the
covariant derivatives
\begin{equation}\label{salg}
i\nabla_{\alpha\dot\alpha}=[\nabla_\alpha, \star
\bar{\nabla}_{\dot\alpha}], \quad [\bar{\nabla}_{\dot\alpha},
\star \nabla_{\beta\dot\beta}]=\epsilon_{\dot\alpha\dot\beta} W_\beta,
\quad [\nabla_\alpha,\star
\nabla_{\beta\dot\beta}]=\epsilon_{\alpha\beta} \bar{W}_{\dot\beta}
\end{equation}
$$
[\nabla_{\alpha\dot\alpha}, \star
\nabla_{\beta\dot\beta}]=-i(\epsilon_{\dot\alpha\dot\beta}f_{\alpha\beta}+
\epsilon_{\alpha\beta}\bar{f}_{\dot\alpha\dot\beta})~.
$$
The
superfields $W_\alpha, \bar{W}_{\dot\alpha}$ satisfy the Bianchi's
identities $\nabla^\alpha \star W_\alpha
+\bar{\nabla}^{\dot\alpha}\star \bar{W}_{\dot\alpha}=0$.
We pay attention that the superstrengths $W_{\alpha}$ and
$\bar{W}_{\dot{\alpha}}$ include the parameter of
nonanticommutativity ${\cal C}^{\alpha\beta}$ by definition.

Classic action for ${\cal N}=1/2$ SYM theory is written using superfield
strengths $W_\alpha$ and $\bar{W}_{\dot\alpha}$ as follows
\begin{equation}\label{clasact}
S= \frac{1}{2g^2}\int d^6z\, {\rm tr}W^\alpha W_\alpha +
\frac{1}{2g^2}\int d^6\bar{z}\, {\rm tr}\bar{W}^{\dot\alpha}
\bar{W}_{\dot\alpha}~.
\end{equation}
 One
can check that the action (\ref{clasact}) can be written in the
form \cite{S}
\begin{equation}
\int d^2\theta \,{\rm tr}\,W^2 =\int d^2\theta \, {\rm
tr}W^2|_{{\cal C}=0} +\int d^4x \,{\rm tr}(-i {\cal
C}^{\alpha\beta}f_{\alpha\beta}\bar{\lambda}^2 -\frac{1}{2}{\cal C
}^2 (\bar{\lambda}^2)^2)~.
\end{equation}
Here $W|_{{\cal C}=0}$ is the superfield strength of the
conventional SYM theory.

Variation of the classic action
$$
\delta S=
\frac{i}{g^2}{\rm tr} \int d^8z\, \left[({\rm e}^{-V}_\star \star
D^\alpha {\rm e}^V_\star ), ({\rm e}^{-V}_\star \star \delta {\rm
e}^V_\star)\right]\star W_\alpha =-\frac{i}{g^2}\int d^8z\,
(\Delta V)_\star \star (\nabla^\alpha \star W_\alpha)~,
$$
leads to classical equations of motion in superfield form
\begin{equation}
\nabla^\alpha \star W_\alpha=0~.
\end{equation}

The dynamics of chiral scalar matter in the (anti)fundamental
representation of the gauge group minimally coupled to the gauge
field is described by the  action
\begin{equation}\label{fund}
S =\int d^8z\,  \bar{\Phi}_f \star {\rm e}^V_{\star} \star  \Phi_f
+\int d^8z \,\tilde{\Phi}_{\bar{f}} \star \,{\rm e}^{-V}_{\star}
\star \,\bar{\tilde{\Phi}}_{\bar{f}} + \int d^6z\, {\cal W}_\star
(\Phi_f, \tilde{\Phi}_{\bar{f}} )+\int d^6\bar{z}\, \bar{\cal
W}_\star (\bar{\Phi}_f, \bar{\tilde{\Phi}}_{\bar{f}})~,
\end{equation}
where $\Phi, \tilde{\Phi}$ are chiral superfields and $\bar{\Phi},
\bar{\tilde{\Phi}}$ antichiral superfields. The kinetic action for
the adjoint representation of the matter fields is
\begin{equation} S =\int
d^8z\, {\rm Tr}({\rm e}^{-V}_{\star}\star \bar{\Phi}\star {\rm
e}^V_{\star} \star  \Phi)~.
\end{equation}
The component field redefinition analogous to the Seiberg-Witten map in
noncommutative non-supersymmetric theories, such that these
fields transform canonically under the gauge transformation was
found in \cite{S}.

In order to formulate the superfield background field method
\cite{11}, \cite{12} for NAC SYM theories we have to perform the
background-quantum splitting ${\rm e}^V_\star \rightarrow {\rm
e}^\Omega_\star \star e^v_\star$ (or ${\rm e}^V_\star \rightarrow
 {\rm e}^v_\star \star {\rm e}^{\bar{\Omega}}_\star$) where $\Omega, (\bar{\Omega})$, $v$ are the
background and quantum superpotentials respectively. Then we have
to write the covariant derivatives in gauge-(anti)chiral
representation as $\nabla_\alpha= {\rm e}^{-v}_\star \star {\bf
\nabla}_\alpha \star {\rm e}^v_\star,
\bar{\nabla}_{\dot\alpha}=\bar{D}_{\dot\alpha}$ ($\nabla_\alpha
=D_\alpha, \bar{\nabla}_{\dot\alpha}={\rm e}^v_\star \star
\bar{\bf \nabla}_{\dot\alpha} \star {\rm e}^{-v}_\star$) with the
standard transformation rules in respect to two gauge
transformations types (quantum and background). The covariantly
(anti)chiral superfields ${\bf \nabla}_\alpha ({\rm
e}^{-{\Omega}}_\star \star\bar{\Phi} ) =\bar{\bf
\nabla}_{\dot\alpha}( {\rm e}^{\bar{\Omega}}_\star \star\Phi) =0$
are splitted linearly into background and a quantum parts.
Background field quantization consists in use of gauge fixing
which explicitly breaks the quantum gauge invariance while
preserves manifest background gauge invariance. The procedure in
NAC case is analogous to conventional one \cite{11}, \cite{12} and
means a replacement the point-products of superfields with the
star-products.

In next sections we consider the effective action induced
by the quantum matter and gauge fields on a special background of $U(1)$
vector multiplet superfield.

\section{The gauge invariant effective action
induced by the matter in the fundamental representation}

We consider the theory with action (\ref{fund}) where the
superfields $\tilde{\Phi}$ are absent. For one-loop calculations
we have to find a quadratic over quantum fields part of the
classical action. After background-quantum splitting defined earlier
one can obtain
\begin{equation}
 S_{(2)} = \frac{1}{2}\int d^8z\; (\bar{\Phi}^T_c, \Phi^T_c)\star
\hat{H}_\star \star \pmatrix{\Phi_c \cr\bar{\Phi}_c} ,\quad
\hat{H}_\star= \pmatrix{ \nabla^2\star
\bar{\nabla}^2&\bar{m}\nabla^2\cr
m\bar{\nabla}^2&\bar{\nabla}^2\star\nabla^2}~,
\end{equation}
where the `masses' are $m={\cal W}_{\Phi\Phi}''(\Phi),
\bar{m}=\bar{\cal W}_{\bar{\Phi}\bar{\Phi}}''(\bar{\Phi}).$

The one-loop correction to the effective action is formally given
by the expression
\begin{equation}\label{one-zeta}
i\Gamma^{(1)}=-\ln \mbox{Det}\hat{H}_\star =-\mbox{Tr} \ln
\hat{H}_{\star}= \zeta' (0)~,
\end{equation}
where $\zeta'(0|H) = \zeta'(\epsilon|H)|_{\epsilon = 0}$ and the
zeta-function  is defined as follows
\begin{equation}\label{zeta}
\zeta(\epsilon|H)=\frac{1}{\Gamma(\epsilon)}\int_0^{\infty} ds
s^{\epsilon-1}\, \mbox{Tr}( {\rm e}^{s\hat{H}_\star}_\star).
\end{equation}
Here $\mbox{Tr}$ is the superspace functional trace $\mbox{Tr}A =
\int d^{8}z A(z,z') \delta^{8}(z-z')|_{z'=z} $ and ${\rm
e}^{s\hat{H}_\star}_\star=1+s\hat{H}_\star+\frac{s^2}{2}\hat{H}_\star
\star \hat{H}_\star+ ... $ Further we follows the procedure
proposed in \cite{16}. Evaluating the effective action on the base
of proper-time techniques consists in two steps: calculation of
the heat kernel, finding the trace and then it renormalization.

First of all we obtain the useful representation of the
$\zeta$-function (\ref{zeta}). Separating the diagonal and
non-diagonal parts of the operator $\hat{H}_\star$ in (\ref{zeta})
we rewrite the $\zeta$-function as
\begin{equation}\label{dz}
\zeta(\epsilon|H) =\frac{1}{\Gamma(\epsilon)}\int^{\infty}_0 \,ds
\cdot s^{\epsilon-1}~ \mbox{Tr}\left( {\rm e}_{\star}^{s\pmatrix{
0&\bar{m}\nabla^2\cr m\bar{\nabla}^2&0}}{\rm e}_{\star}^{s\pmatrix{
\nabla^2\star \bar{\nabla}^2&0\cr0
&\bar{\nabla}^2\star\nabla^2}}\right)
\end{equation}
$$
 =\int \,d^8 z \int^{\infty}_0 \frac{ds}{\Gamma(\epsilon)}
s^{\epsilon-1} \sum^{\infty}_{n=0} \frac{s^{2n}}{(2n)!}
(m\bar{m})^n \frac{d^n}{ds^n} {\rm e}_{\star}^{{s}\nabla^2
\star\bar{\nabla}^2} \delta^{8}(z-z')|_{z'=z}  + (\nabla^2
\leftrightarrow \bar{\nabla}^2)
$$
$$
=\int \,d^6 z \int^{\infty}_0 \frac{ds}{\Gamma(\epsilon)}
s^{\epsilon-1} \sum^{\infty}_{n=0} \frac{s^{2n}}{(2n)!}
(m\bar{m})^n \frac{d^n}{ds^n} {\rm e}_{\star}^{{s}\bar{\nabla}^2
\star{\nabla}^2}\star \bar{\nabla}^2 \delta^{8}(z-z')|_{z'=z} +
\int d^6 \bar{z} (\nabla^2 \leftrightarrow \bar{\nabla}^2)~.
$$
Here we used property $d^8 z=d^6 z\bar{\nabla}^2=d^6
\bar{z}\nabla^2$ and fulfilled integration by parts. Also we suggest
that the `masses' are slowly varying. The `mass' dependence in
(\ref{dz}) is accompanied by derivatives of the (anti)chiral
kernels. It is convenient to separate these derivatives from
`masses'. It can be done by repeated integration by parts. Direct
calculation with keeping in mind a limit $\epsilon \rightarrow 0$
leads to the following representation of $\zeta$-function
\begin{equation}\label{zet}
\zeta(\epsilon|H)=\frac{1}{\Gamma(\epsilon)} \int_0^{\infty}
\,ds\cdot s^{\epsilon-1} {\rm e}^{-m\bar{m}s} \left(K_{+}(s) +
K_{-}(s)\right)~,
\end{equation}
with the chiral and antichiral heat kernel traces defined as
\begin{equation}\label{kerns+}
K_{+}(s)= \frac{1}{2}\int d^6z {\rm e}^{s \Box_{+}}_\star \star
\bar{\nabla}^2 \delta^8(z-z')|_{z=z'}~, K_{-}(s)= \frac{1}{2}\int
d^6\bar{z}\, {\rm e}^{s \Box_{-}}_\star \star {\nabla}^2
\delta^8(z-z')|_{z=z'}~.
\end{equation}
In above expressions we used the Laplace-type operators acting in
the space of covariantly (anti)chiral superfields ($\Box_{+} \Phi
=\bar{\nabla}^2 \star \nabla^2 \star \Phi$, $\Box_{-} \bar{\Phi}
={\nabla}^2 \star \bar{\nabla}^2 \star \bar{\Phi}$)
\begin{equation}
\Box_{+} =\Box_\star -iW^\alpha \star \nabla_\alpha
-\frac{i}{2}(\nabla^\alpha \star W_\alpha), \quad \Box_{-}
=\Box_\star -i\bar{W}^{\dot\alpha} \star \bar{\nabla}_{\dot\alpha}
-\frac{i}{2}(\bar{\nabla}^{\dot\alpha} \star
\bar{W}_{\dot\alpha})~,
\end{equation}
where $\Box_\star =\frac{1}{2}\nabla^{\alpha\dot\alpha} \star
\nabla_{\alpha\dot\alpha}~.$ Relations (\ref{one-zeta}, \ref{zet},
\ref{kerns+}) define the gauge invariant one-loop effective action
for the theory under consideration.

Further we will consider an approximation of a covariantly
constant on-shell background vector multiplet where the  of
effective action can be carried up the very end and a result can
be expressed in a closed form. Such a background is defined as
follows
\begin{equation}\label{onshell}
\nabla_{\alpha\dot\alpha} \star
W_\beta=0, \quad \nabla^\alpha \star W_\alpha=0.
\end{equation}
The one-loop contribution to the effective action is found
using the methods formulated in the Refs
\cite{18}, \cite{16} and taking into account the property (\ref{main})
of the star-product. As a result one gets
\begin{equation}\label{g1matter}
\begin{array}{l}
\displaystyle \Gamma^{(1)}=-\frac{1}{(4\pi)^2} \int d^6z \,W^2
\ln\frac{m}{\Lambda} + c.c. \\
\displaystyle+\frac{1}{(4\pi)^2}\int^\infty_0 d s \cdot s \,{\rm
e}^{-m\bar{m}s} \int d^8z \,W^2\star \bar{W}^2\star \zeta_\star
(s {\cal N},s\bar{\cal N})\\
\end{array}
\end{equation}
where ${\cal
N}_\alpha^\beta =D_\alpha W^\beta \footnote{${\cal
N}_\alpha^\beta{\cal N}^\delta_\beta =\delta^\delta_\alpha D^2 W^2
=\delta^\delta_\alpha {\cal N}^2 $}$, $\bar{\cal
N}_{\dot\alpha}^{\dot\beta} =\bar{D}_{\dot\alpha}
\bar{W}^{\dot\beta}$ and function $\zeta(x,y)$ has been introduced
in \cite{18}
\begin{equation}\label{zetastar} \zeta_\star
(x,y)=\frac{y^2 \star (\cos_\star x -1)-x^2 \star(\cos_\star
y-1)}{x^2\star y^2\star (\cos_\star x -\cos_\star y)}~.
\end{equation}

Thus, in the theory under consideration with the operators
$\Box_{\pm}$ including only left star-products, the heat trace
expansion and hence the effective action is completely defined by
ones in conventional superfield theory, so the only difference is
presence of $\star$-products in final results instead of ordinary
products.

\section{Heat kernel and the effective action in the NAC SYM theory}

In this section we consider the one-loop contributions to the
effective action of the gauge fields and ghosts. We study a theory
with $SU(2)$ gauge group broken to $U(1)$ and assume that the
background is described by on-shell Abelian superfield
(\ref{onshell}).

\subsection{Features of the background-quantum splitting
in the NAC SYM theory}

We describe a structure of the background-quantum splitting in the NAC
SYM model and point out a role of the operators $T^{\star}_{c,s}$
(\ref{t}) for the fields in an adjoint representation.

Using the background-quantum splitting $\nabla \rightarrow {\rm e}^{-v}
\nabla {\rm e}^v$ and the standard gauge fixing function $\chi =\nabla^2 v$
leads to the following
quadratic part of ${\cal N}=1/2$ SYM action with $SU(2)$ gauge group for the quantum
superfield $v$
(see \cite{11} for some details in conventional superfield theory)
\begin{equation}\label{quadsym}
S_{gauge+gf}^{(2)}=  -\frac{1}{2g^2} \mbox{tr} \int d^8 z\, v \star
H_{\star}^{(v)}\star v~
\end{equation}
where the operator $H_{\star}^{v}$ has the form
\begin{equation}\label{H}
H_{\star}^{(v)}= \Box_{\star}-i\{W^\alpha, \star \mathbb{\nabla}_\alpha\}
-i\{\bar{W}^{\dot\alpha}, \star
\bar{\mathbb{\nabla}}_{\dot\alpha}\}~
\end{equation}
Similarly, the quadratic part of the ghost action has the form
\begin{equation}
S_{ghosts}^{(2)}=\mbox{tr} \int d^8z \,(\bar{c}' c -c'\bar{c}
+\bar{b} b)~.
\end{equation}
where all ghost superfields are background
covariantly (anti)chiral. The quantum superfield $v$ and
the ghost superfields $c$, $c'$, $b$  belong to Lie algebra
$su(2)$. It means that $v=v^a \tau_a$ and the same true for the ghost superfields.
Here $\tau_a=\frac{1}{\sqrt{2}}\sigma_a$ are the generators of $su(2)$
algebra satisfying the relations $[\tau_a,\tau_b]=i\sqrt{2} \epsilon_{abc}\tau_c,
\mbox{tr}(\tau_a \tau_b)=\delta_{ab}$.

Structure of the operator $H_{\star}^{(v)}$ (\ref{H}) for the
background belonging to the Abelian subgroup $U(1)$ (in this case
$W=W^3\tau_3$) can be simplified because of the property of the star-operator $T^{*}_{c}$.
The operator $H_{\star}^{(v)}$ (\ref{H}) includes
the term ${W^\alpha \star \mathbb{\nabla}_{\alpha}}$. We write
${\nabla}_{\alpha}= D_{\alpha}- i\Gamma_{\alpha}$ and consider the
term in $H_{\star}^{(v)}$ containing only $D_{\alpha}$
\begin{equation}
v\{W,\star Dv\}=\tau^a \tau^c \tau^b (v^a W^c \star D v^b)+\tau^a
\tau^b \tau^c (v^a D v^b \star W^c )
\end{equation}
$$
= \frac{1}{2}\tau^a [\tau^c, \tau^b] v^a [W^c ,\star D
v^b]+\frac{1}{2}\tau^a \{\tau^c, \tau^b\} v^a \{W^c ,\star D
v^b\}~.
$$
As a result one gets
\begin{equation}\label{T}
{\rm tr}\left( v \{W^\alpha \star D_\alpha v\}\right) ={\rm
tr}\left(\tau^a[\tau^c,\tau^b]\right)v^a W^\alpha_c T^\star_c
D_\alpha v^b + {\rm tr}\left(\tau^a\{\tau^c,\tau^b\}\right)v^a
W^\alpha_c T^\star_s D_\alpha v^b~,
\end{equation}
where we have used the definitions $T^\star_{c(s)}=\frac{1}{2}(\star
\pm \star^{-1})$ to rewrite the $[W, \star Dv]$ and $\{W, \star
Dv\}$. The last term in (\ref{T}) is equal to zero because of
$\mbox{tr}(\tau^a\{\tau^c,\tau^b\})=0$, while in the first term only
component $W_3$ survives. After redefinition of the gauge field
components $v^{1}$ and $v^{2}$ as follows
$\chi=\frac{1}{\sqrt{2}}(v^1+iv^2),\quad
\tilde{\chi}=\frac{1}{\sqrt{2}}(v^1-iv^2) $ the first term can be
rewritten in the form \begin{equation}\label{three} \sqrt{2}\chi
W_3^\alpha T^\star_c D_\alpha
\tilde{\chi}-\sqrt{2}\tilde{\chi}W_3^\alpha T^\star_c D_\alpha
\chi~,
\end{equation}
while the $v^{3}$ component of the quantum superfield
$v=v^{a}\tau_{a}$  do not interact with the background and totally
decouple. Above we have analyzed a contribution of the operator
$D_{\alpha}$ from operator ${\nabla}_{\alpha}$ in (\ref{H}). Now
let us consider a contribution of another term $\Gamma_{\alpha}$
which forms together with $D_{\alpha}$ the supercovariant
derivative ${\nabla}_{\alpha}= D_{\alpha}- i\Gamma_{\alpha}$. Its
contribution to ${\rm tr} (v \star H_{\star}^{(v)}\star v)$ is
given by
\begin{equation}
{\rm tr}\left(v\star\{W \star[\Gamma,\star
v]\}\right)=\end{equation} $$ {\rm tr}\left(
\frac{1}{4}[\tau_a,\tau_c]\{\tau_d,\tau_b\}\right) v^a [W^c, \star
[\Gamma^d,\star v^b]] + {\rm tr}\left(\frac{1}{2}\{\tau_a,\tau_c
\}\frac{1}{2}\{\tau_d,\tau_b \}\right)v^a\{W^c, \star [\Gamma^d,
\star v^b]\} $$ $$ +{\rm tr}\left(\frac{1}{2}\{\tau_a,\tau_c
\}\frac{1}{2}[\tau_d,\tau_b ]\right)v^a\{W^c, \star \{\Gamma^d,
\star v^b\}\} + {\rm tr}\left(\frac{1}{2}[\tau_a,\tau_c
]\frac{1}{2}[\tau_d,\tau_b ]\right)v^a[W^c, \star \{\Gamma^d,
\star v^b\}]~.
$$
It is easy to see that the first and the third terms are equal to
zero due to the trace properties of $\tau$-matrices. The second
term is proportional to  $v^3 W^\alpha T^\star_s \Gamma_\alpha
T^\star_s v^3$ and will not give a contribution to the effective
action, because of the property (\ref{sinh}) for star-operator
$T^\star_s$. The last term together with (\ref{three}) can be
rewritten as $\chi [-iW^\alpha_3 T^\star_c
\nabla_\alpha]\tilde{\chi}$, where now $\nabla_\alpha=D_\alpha
-i\Gamma_\alpha T^\star_c$. Appearance of $T^{\star}_{c}$ was
stipulated by the adjoint representation.

As a result we found that the contributions of $v^{3}$- component
of the quantum gauge multiplet totally decouple. Moreover,
according to the property of $T^\star_s$-product, given in
the Section 2, we see that theirs contributions the one-loop
effective action are absent. Non-trivial contribution to the
effective action is generated by the components $v^{1}$ and
$v^{2}$ or by their linear combinations $\chi$ and $\tilde{\chi}$.
We want to emphasize that the action for the $\chi, \tilde{\chi}$
corresponds to non-Abelian superfield model where the
star-operator $T^{\star}_{c}$  plays the role of an internal
symmetry generator including whole star-structure of the initial
theory. Further we study a construction of the heat kernel and the
effective action of the theory under consideration generalizing
the techniques \cite{13} for NAC superspace.

\subsection{The heat kernels on the covariantly constant background}

Above we have shown that the second variational derivative of the
action in sector of the superfields $\chi, \tilde{\chi}$ has the
form $S^{(2)}_{gauge +FG}=\int d^8z \,\chi H^\chi_{\tilde{\star}}
\tilde{\chi}$ where the operator $H^\chi_{\tilde{\star}}$ defined
as
\begin{equation}\label{hnab}
H^\chi_{\tilde{\star}} =\Box_{\tilde{\star}}
-iW^\alpha_{\tilde{\star}} \nabla_{\tilde{\star}\, \alpha}
-i\bar{W}^{\dot\alpha}_{\tilde{\star}}
\bar{\nabla}_{\tilde{\star}\, \dot\alpha}~,
\end{equation}
and the notations $W_{\tilde{\star}}=W T^\star_c$ and
$\nabla_{\tilde{\star}}$ for $D-i\Gamma T^\star_c$ were used. We
define the Green function $G(z,z')$ of the operator
$H^{\chi}_{\tilde{\star}}$ by the equation
$H^{\chi}_{\tilde{\star}}G(z,z') = -\delta^8(z-z')$. Then one
introduces the heat kernel $K_{\chi}(z,z'|s)$ associated with this
Green function as $G(z,z')= \int^\infty_0 ds \,K(z,z'|s)\, {\rm
e}^{-\varepsilon s}|_{\varepsilon \rightarrow +0}.$ It means that
formally $K_\chi(z,z'|s)={\rm e}^{sH^\chi_{\tilde{\star}}}
\delta^8(z-z')$. The one-loop contribution of the gauge
superfields to the effective action is proportional to ${\rm Tr}
(K_{\chi})$ and gauge invariant due to the gauge transformation
law $$K_{\chi}(z,z') \rightarrow {\rm
e}^{i\Lambda(z)}_{\tilde{\star}} K_{\chi}(z,z'|s) {\rm
e}^{-i\Lambda(z')}_{\tilde{\star}}.$$ We rewrite the kernel
$K_{\chi}$ in the form
\begin{equation}\label{kchi}
K_\chi (z,z'|s) = {\rm e}^{s(\Box_{\tilde{\star}}
-iW^\alpha_{\tilde{\star}} \nabla_{{\tilde{\star}}\, \alpha}
-i\bar{W}^{\dot\alpha}_{\tilde{\star}}\bar{\nabla}_{{\tilde{\star}}\,
\dot\alpha})} \times (\delta^8(z-z')I(z,z'))~,
\end{equation}
where bi-scalar $I(z,z')$ satisfies the equation
$\zeta_A(z,z')\nabla_{\tilde{\star}}^A I(z,z')=0$ and boundary
condition $I(z,z)=1$. Further we will use the techniques developed in
\cite{13} adopting it to the NAC superspace.

In order to calculate (\ref{kchi}) we, first of all, write the
operator $H^\chi_{\tilde{\star}}$ (\ref{hnab}) as follows
$H^\chi_{\tilde{\star}} = \Box_{\tilde{\star}} + V$ where $V =
 -iW^\alpha_{\tilde{\star}} \nabla_{\tilde{\star}\, \alpha}
-i\bar{W}^{\dot\alpha}_{\tilde{\star}}
\bar{\nabla}_{\tilde{\star}\, \dot\alpha}$ and decompose the operator
${\rm e}^{sH^\chi_{\tilde{\star}}}$ as
\begin{equation}\label{expbm}
{\rm e}^{s(\Box_{\tilde{\star}} +V)} = \cdots \times{\rm
e}^{+\frac{s^3}{2}[V,[\Box_{\tilde{\star}}, V]]
+\frac{s^3}{6}[\Box_{\tilde{\star}},[\Box_{\tilde{\star}},V]]}{\rm
e}^{\frac{s^2}{2}[\Box_{\tilde{\star}}, V]}{\rm e}^{s V} {\rm
e}^{s \Box_{\tilde{\star}}}~.
\end{equation}
It follows from (\ref{salg}) that for $\nabla_{{\tilde{\star}}\,
\alpha\dot\alpha}W_{{\tilde{\star}}\, \beta}=0$ the first
commutator becomes
\begin{equation} [\Box_{\tilde{\star}}, V]=(iW_\alpha T^\star_c
\bar{W}_{\dot\alpha}+i\bar{W}_{\dot\alpha}T^\star_c
W_{\alpha})T^\star_c
\nabla^{\alpha\dot\alpha}_{\tilde{\star}}=0~,\end{equation}
because of the property $W_\alpha(\star
+\star^{-1})\bar{W}_{\dot\alpha}=
-\bar{W}_{\dot\alpha}(\star^{-1}+\star\,)W_{\alpha}$ which is
valid for the chosen Abelian background. This identity leads to
convenient factorization of the kernel in the form (see \cite{13}
for details in conventional superfield theory)
\begin{equation}\label{kern}
K_\chi(z,z'|s) =U_{\tilde{\star}}(s)\,\tilde{K}(\zeta|s)
\,\zeta^2 \bar{\zeta}^2\, I(z,z'), \quad U_{\tilde{\star}}(s)={\rm
e}^{-is(W^\alpha_{\tilde{\star}} \nabla_{{\tilde{\star}}\, \alpha}
+ \bar{W}^{\dot\alpha}_{\tilde{\star}}
\bar{\nabla}_{{\tilde{\star}}\, \dot\alpha})}~.
\end{equation}
Here we have used the chiral basis with coordinates
$(y^{\alpha\dot{\alpha}}, \theta^{\alpha}, \theta^{\dot{\alpha}})$
for calculations and the presentation
$\delta^8(z-z')=\delta^4(\zeta^{\alpha\dot\alpha})\zeta^2\bar{\zeta}^2$.
The translation invariant interval components $\zeta^{A}(z, z')$
in the chiral basis are defined by
$$
\zeta^A = (\zeta^{\alpha\dot\alpha}, \zeta^{\alpha},
\bar{\zeta}^{\dot\alpha}) =
((y-y')^{\alpha\dot\alpha}-i(\theta-\theta')^\alpha
\bar{\theta}^{' \dot\alpha}, (\theta-\theta')^\alpha,
(\bar{\theta}-\bar{\theta})^{\dot\alpha})~.
$$

The Schwinger type heat kernel $\tilde{K}(\zeta|s)$ in
(\ref{kern}) can be calculated by the various methods and it is
well known
\begin{equation}\label{sch}
\tilde{K}(\zeta|s)=\frac{i}{(4\pi s)^2}
\exp({-\frac{1}{2}\mbox{tr}\ln_{\tilde{\star}}\frac{\sin_{\tilde{\star}}
sF/2}{sF/2}}){\rm e}^{-\frac{1}{2 s}\zeta(\frac{sF}{2}
\cot_{\tilde{\star}} \frac{sF}{2}) \zeta}~,
\end{equation}
where
$F^{\beta\dot\beta}_{\alpha\dot\alpha}=\delta^{\dot\beta}_{\dot\alpha}f^\beta_\alpha
+ \delta^{\beta}_{\alpha}\bar{f}^{\dot\beta}_{\dot\alpha} $ and
$f^{\beta}_{\alpha}, \bar{f}^{\dot{\beta}}_{\dot{\alpha}}$ are the
spinor components of the Abelian strengths $F_{mn}$\footnote{Also
we point out the useful equation $\nabla_{{\tilde{\star}}\,
\alpha\dot\alpha}\tilde{K} +\left(\frac{i F}{{\rm
e}_{\tilde{\star}}^{isF}-1}\right)^{\beta\dot\beta}_{
\alpha\dot\alpha}\zeta_{\beta\dot\beta}{\tilde{\star}} \tilde{K}
=0$, which allows us to get any order derivatives of the Schwinger
type kernel.}. The contraction over indices in the right exponent
goes only for $\zeta^{\alpha\dot\alpha}$ components. The functions
$f_{\tilde{\star}}(x)$ are given as an expansion
$f_{\tilde{\star}}(x)= \sum_{n=0}^\infty
\frac{f^{(n)}}{n!}(xT^\star_c(x T^\star_c(... T^\star_c x)))$.

Next step of calculations is obtaining the action of the operator
$U_{\tilde{\star}}(s)$ in (\ref{kern}). This operator contains covariant
derivatives which act on the interval components
$U_{\tilde{\star}}(s)\zeta^A = \zeta^A(s) U_{\tilde{\star}}(s)$.
Hence we should consider the adjoint action of $U$ on $\zeta^A$.
Introducing the notation ${\cal N}_{{\tilde{\star}}\,
\alpha}^\beta=\nabla_{{\tilde{\star}}\, \alpha}
W^\beta_{\tilde{\star}}$, $\bar{\cal N}_{{\tilde{\star}}\,
\dot\alpha}^{\dot\beta}=\bar{\nabla}_{{\tilde{\star}}\,
\dot\alpha} \bar{W}_{\tilde{\star}}^{\dot\beta}$ we have for
adjoint action $U$ on the interval components
\begin{equation}\label{zetaint}
{\zeta}^{\alpha}(s) ={\zeta}^{\alpha} +W^\delta_{\tilde{\star}}
\left(\frac{{\rm e}^{-is{\cal N}_{\tilde{\star}}}-1}{{\cal
N}_{\tilde{\star}}}\right)_\delta^\alpha ,\quad
\bar{\zeta}^{\dot\alpha}(s) =\bar{\zeta}^{\dot\alpha}
+\bar{W}^{\dot\delta}_{\tilde{\star}}\left(\frac{{\rm
e}^{-is\bar{\cal N}_{\tilde{\star}}}-1}{\bar{\cal
N}_{\tilde{\star}}}\right)_{\dot\delta}^{\dot\alpha}~,
\end{equation}
\begin{equation}
\zeta^{\alpha\dot\alpha}(s)=\zeta^{\alpha\dot\alpha} +\int_0^s d
\tau W^{\alpha}_{\tilde{\star}}(\tau)
\bar{\zeta}^{\dot\alpha}(\tau)~,
\end{equation}
where
$W^\alpha(s)_{\tilde{\star}}=W^\beta_{\tilde{\star}} (e^{-is{\cal
N}_{\tilde{\star}}})_\beta^\alpha$.

Next step is calculation of $U_{\tilde{\star}} I(z,z')$ in
(\ref{kern}). To do that we write a differential equation
\begin{equation}\label{eq}
i\frac{d}{d s}U_{\tilde{\star}}(s) I(z,z') =
U_{\tilde{\star}}(s)(W_{\tilde{\star}}\nabla_{\tilde{\star}}
+\bar{W}_{\tilde{\star}}\bar{\nabla}_{\tilde{\star}})U_{\tilde{\star}}^{-1}(s)
U_{\tilde{\star}}(s)I(z,z')
\end{equation}
and solve it. Thus, we should construct the operators
$\nabla_{\tilde{\star}A}(s)$ and act on $I(z,z')$. We pay
attention that the procedure of calculations, we discuss here,
preserves manifest gauge invariance. Therefore, to simplify the
calculations, we can impose any appropriate gauge on background
superfield. The treatment with $I(z,z')$ are very much simplified
under conditions $I(z,z')=1$ which is equivalent to the
Fock-Schwinger gauge $\zeta^{A}\star \Gamma_{\tilde{\star}A} = 0$
or $\nabla_{\tilde{\star}A}I(z,z') =
-iI(z,z')\Gamma_{\tilde{\star}A}$ (see details in \cite{13} for
conventional superspace theories).

If in the chiral basis we have a supercovariant derivative
$\nabla'_{\tilde{\star}A}$ in the point $z'$, then the
supercovariant derivative $\nabla_{\tilde{\star}}$ satisfying the
Fock-Schwinger gauge in point $z$ has the form
\begin{equation}\label{chir}
\nabla_{{\tilde{\star}}\, A} ={\rm e}^{+i y^{' m} Q_m +i\theta^{'}
Q}{\rm e}^{+i\bar{\theta}'\bar{Q}}{\rm
e}^{+y^{\alpha\dot\alpha}\nabla'_{{\tilde{\star}}\,
\alpha\dot\alpha}+\theta^\alpha \nabla'_{{\tilde{\star}}\, \alpha}}
{\rm e}^{+\bar{\theta}^{\dot\beta} \bar{\nabla}'_{{\tilde{\star}}\,
\dot\beta}}{\rm e}^{-\bar{\theta}^{'
\dot\alpha}\bar{D}_{\dot\alpha}}{\rm e}^{-y^{'
\beta\dot\beta}\partial_{\beta\dot\beta} -\theta^{'\alpha}
D_\alpha}\times
\end{equation}
$$(\nabla'_{{\tilde{\star}}\,A}) \times \quad {\rm e}^{y^{'
\beta\dot\beta}\partial_{\beta\dot\beta} +\theta^{'\alpha}
D_\alpha} {\rm e}^{\bar{\theta}^{'
\dot\alpha}\bar{D}_{\dot\alpha}} {\rm
e}^{-\bar{\theta}^{\dot\beta} \bar{\nabla}'_{{\tilde{\star}}\,
\dot\beta}} {\rm
e}^{-y^{\alpha\dot\alpha}\nabla'_{{\tilde{\star}}\,
\alpha\dot\alpha}-\theta^\alpha \nabla'_{{\tilde{\star}}\,
\alpha}} {\rm e}^{-i\bar{\theta}'\bar{Q}} {\rm e}^{-i y^{' m} Q_m
-i\theta^{'} Q}~ $$ The relation (\ref{chir}) leads to explicit
expressions for the connections in the Fock-Schwinger gauge:
\begin{equation}\label{confs}
\bar{\nabla}_{{\tilde{\star}}\, \dot\beta}-\bar{D}_{\dot\beta}=
-i\Gamma_{{\tilde{\star}}\, \dot\beta}=0~,
\end{equation}
$$
\nabla_{{\tilde{\star}}\, \beta}-D_\beta
=-i\Gamma_{{\tilde{\star}}\, \beta}
=\frac{1}{2}\zeta^{\dot\beta}_{\beta L} \bar{W}'_{{\tilde{\star}}\,
\dot\beta}-\frac{i}{2}\zeta_\beta
(\bar{\zeta}^{\dot\beta}\bar{W}'_{{\tilde{\star}}\, \dot\beta})
-\frac{1}{2}\zeta^{\alpha\dot\alpha}_L
\bar{\zeta}^{\dot\beta}F'_{{\tilde{\star}}\, \beta\dot\beta,
\alpha\dot\alpha} +i\bar{\zeta}^2(W'_{{\tilde{\star}}\, \beta}
+\zeta^\alpha(\nabla'_\alpha W'_{{\tilde{\star}}\, \beta}))~,
$$
$$
\nabla_{{\tilde{\star}}\,
\beta\dot\beta}-\partial_{\beta\dot\beta}=-i\Gamma_{{\tilde{\star}}\,
\beta\dot\beta}= -\frac{i}{2}\zeta^{\alpha\dot\alpha}_L
F'_{{\tilde{\star}}\, \alpha\dot\alpha,
\beta\dot\beta}+\bar{\zeta}_{\dot\beta}W'_{{\tilde{\star}}\, \beta}
+\frac{1}{2}\zeta_\beta \bar{W}'_{{\tilde{\star}}\, \dot\beta}
+\bar{\zeta}_{\dot\beta}\zeta^\alpha (\nabla'_{{\tilde{\star}}\,
\alpha} W'_{{\tilde{\star}}\, \beta}).
$$
First of these relations is the consequence of the
supercovariant derivative forms in the chiral basis (\ref{nabla}). Using
(\ref{confs}) one can find the solution of the equation (\ref{eq})
in the form
\begin{equation}\label{ui}
U_{\tilde{\star}}(s) I(z,z')=\exp_{\tilde{\star}}\left(i\int^s_0
d\tau (\frac{1}{2}W_{{\tilde{\star}}\,
\beta}(\tau)\zeta^{\beta\dot\beta}(\tau)\bar{W}_{{\tilde{\star}}\,
\dot\beta}'
+\frac{i}{2}W_{\tilde{\star}}^\beta(\tau)\zeta_\beta(\tau)
\bar{\zeta}^{\dot\beta}(\tau)\bar{W}_{{\tilde{\star}}\,
\dot\beta}'\right.
\end{equation}
$$ \left.+\frac{1}{2}W_{\tilde{\star}}^\beta(\tau)
\zeta^{\alpha\dot\alpha}(\tau)
\bar{\zeta}^{\dot\beta}(\tau)F_{{\tilde{\star}}\, \beta\dot\beta
\alpha\dot\alpha}' -i\bar{\zeta}^2(\tau)
W_{\tilde{\star}}^\beta(\tau) W_{{\tilde{\star}}\beta}'
-i\bar{\zeta}^2(\tau) W_{\tilde{\star}}^\beta(\tau)
\zeta^\alpha(\tau) f_{{\tilde{\star}}\, \alpha\beta}')\right)~.
$$

Substituting the $\zeta^{A}(s)$ (\ref{zetaint}) and
$U_{\tilde{\star}} I(z,z')$ (\ref{ui}) into (\ref{kern}) and
taking into account (\ref{sch}) one gets finally the kernel
\begin{equation}\label{kerns}
K_\chi(z,z'|s) = \tilde{K}(\zeta|s) \,\zeta^2(s)
\bar{\zeta}^2(s) U_{\tilde{\star}}(s)I(z,z')~.
\end{equation}
determining the effective action.

Now we discuss a structure of kernels corresponding to the ghost or to
any adjoint chiral matter contribution to the effective action.
First of all we point out that the following relations take place
in on-shell Abelian background
\begin{equation}
\bar{\nabla}^2_{\tilde{\star}} {\rm e}^{sH^\chi_{\tilde{\star}}}=
\bar{\nabla}^2_{\tilde{\star}} {\rm e}^{s
\nabla^2_{\tilde{\star}}\bar{\nabla}^2_{\tilde{\star}}}= {\rm e}^{s
\bar{\nabla}^2_{\tilde{\star}}{\nabla}^2_{\tilde{\star}}}\bar{\nabla}^2_{\tilde{\star}}
={\rm e}^{s \Box_{{\tilde{\star}}\,
+}}\bar{\nabla}^2_{\tilde{\star}},
\end{equation}
\begin{equation}
{\nabla}^2_{\tilde{\star}} {\rm e}^{sH^\chi_{\tilde{\star}}}=
{\nabla}^2_{\tilde{\star}} {\rm e}^{s
\bar{\nabla}^2_{\tilde{\star}}{\nabla}^2_{\tilde{\star}}}= {\rm
e}^{s {\nabla}^2_{\tilde{\star}}
{\bar{\nabla}}^2_{\tilde{\star}}}{\nabla}^2_{\tilde{\star}} ={\rm
e}^{s \Box_{{\tilde{\star}}\,-}}{\nabla}^2_{\tilde{\star}}~.
\end{equation}
Let us introduce the chiral and antichiral heat kernels
\begin{equation}\label{+-}
K_{+}(z,z'|s)=\bar{\nabla}^2_{\tilde{\star}} K_\chi
(z,z'|s)=\bar{\nabla}^{' 2}_{\tilde{\star}} K_\chi (z,z'|s),
\end{equation}
$$ K_{-}(z,z'|s)=\nabla^2_{\tilde{\star}}
K_{\chi}(z,z'|s)=\nabla^{' 2}_{\tilde{\star}} K_{\chi}(z,z'|s)~.
$$
The functions $G_{\pm}(z,z')$ defined as $G_{\pm}(z,z')=
\int^\infty_0 ds \,K_{\pm}(z,z'|s)\, {\rm e}^{-\varepsilon
s}|_{\varepsilon \rightarrow +0}$ satisfy the equations
$\Box_{{\tilde{\star}}\, \pm}G_{\pm}(z,z')=-\delta_{\pm}(z,z')$
where $\Box_{{\tilde{\star}}\, \pm}$ are the (anti)chiral
d'Alamber\-tians depending on background superfield in NAC
superspace. It means that the functions (\ref{+-}) are the kernels
associated with the operators $\Box_{{\tilde{\star}}\, \pm}$.
Namely these kernels determine the one-loop contribution to the
effective action from any chiral matter in the adjoint representation.
Since the kernels (\ref{+-}) are from the kernel
$K_{\chi}$ we can substitute (\ref{kerns}) into (\ref{+-}) and
find these kernels
\begin{equation}\label{chker}
K_{+}(z,z'|s) =\bar{\nabla}_{\tilde{\star}}^{' 2}K_\chi =-
\tilde{K}(\zeta|s) \zeta^2(s)  U_{\tilde{\star}}(s)\,I(z,z')
\end{equation}
\begin{equation}\label{achker}
K_{-}(z,z'|s) =\nabla_{\tilde{\star}}^{' 2}K_\chi
=-\tilde{K}(\zeta|s) \, \bar{\zeta}^2(s){\rm
e}_{\tilde{\star}}^{-\frac{1}{2}\zeta^\alpha(s)
\zeta_{\alpha\dot\alpha}(s)\bar{W}_{\tilde{\star}}^{\dot\alpha}(s)}
U_{\tilde{\star}}(s)\,I(z,z')~.
\end{equation}
These relations determine a contribution of the chiral adjoint matter
including ghost superfields into the one-loop effective action.

In next subsection we consider heat traces associated with the heat
kernels (\ref{kerns}, \ref{chker}, \ref{achker}) and, particularly,
coefficient $a_2$ in the Schwinger-De Witt expansion of the
low-energy effective action.

\subsection{Gauge fields and ghosts contribution to the effective
action\\ $SU(2)$ SYM theory}

The one-loop contribution $\Gamma^{(1)}$ to the effective action
for NAC SYM theory is defined with the help of the
$\zeta$-functions $\zeta_{\chi, \pm}(\epsilon)$ corresponding to
the operators $H^{\chi}_{\star}$, $\Box_{{\tilde{\star}}\, \pm}$
respectively
\begin{equation}\label{Gamma}
\Gamma^{(1)}= \Gamma_{\chi}^{(1)}+ \Gamma_{ghosts}^{(1)}.
\end{equation}
Here $\Gamma_{\chi}^{(1)}$ is pure SYM contribution and
$\Gamma_{ghosts}^{(1)}$ is ghost contribution. Each of these
contributions is calculated via function $\zeta'(0)$. In its turn,
the $\zeta$-functions are given by integral representations
\begin{equation} \zeta_{\chi, \pm
}(\epsilon)=\frac{1}{\Gamma(\epsilon)}\int^\infty_0 \frac{d
s}{s^{1-\epsilon}}\mbox{Tr} K_{\chi,\pm}(\frac{s}{\mu^2})
\end{equation}
where $\mu$ is the normalization point and $\mbox{Tr}$ is an
inherent functional trace.

General structure of $K_{\chi,\pm}(z,z'|s)$ was discussed in the
Subsection 5.2 (see (\ref{kerns}, \ref{chker}, \ref{achker})). One
can show that ${\rm Tr}K_\chi(s)$ does not contain the
holomorphic contributions and, hence, the one-loop divergent
contributions to the complete effective action (\ref{Gamma}) is
determined exclusively by the ghosts as in the conventional case
\cite{11}.

To construct the divergent part of the effective action we should
consider the behavior of $K_{\pm}(z,z'|s)|_{z'=z}$ at small $s$.
As usual, the kernel expansion looks like $K_{\pm}(z,z|s) \sim
\frac{1}{s^2} (a_{0}(z,z) + s a_{1}(z,z) + s^2 a_{2}(z,z) +
\ldots)$ The coefficient $a_{2}(z,z)$ is responsible for the
divergences. Exact form of the kernel $K_{\pm}(z,z'|s)$ is given
by (\ref{chker}, \ref{achker}). The only thing we should do is to
study its behavior at small $s$. One can show that the
coefficients $a_{0}(z,z) = 0,$ $a_{1}(z,z) = 0$. The coefficient
$a_{2}(z,z)$ includes the products of some number of $W_{\alpha}$,
some number of superintervals $\zeta^A$ and some number of
star-operators $T^{\star}_{c}$ acting on the intervals and
superstrengths. Using the explicit form (\ref{t}) of the operator
$T^{\star}_{c}$ one can show that the final expression for
$a_{2}(z,z)$ is a sum of one for conventional superfield theory
plus a total derivative with respect to the variable
$\theta^{\alpha}$ which is stipulated by action of the operator
$T^{\star}_{c}$ on the superinterval $\zeta^A$. As a result one
obtains
\begin{equation}\label{div}
\int d^6z \, a_{2}(z,z|s) \sim \int d^6z \,W^\alpha W_\alpha
\end{equation}
It leads to the divergent part of the effective
action in the form
\begin{equation}
\Gamma^{(1)}_{div}=-\frac{3}{2}\cdot\frac{1}{(4\pi)^2}\int d^6 z
W^\alpha W_\alpha \ln\frac{\mu^2}{\Lambda^2}.
\end{equation}
We see that the one-loop divergences on the Abelian background
are analogous to classical action what provides
renormalizability.

The finite parts of the effective action is analyzed by known
methods (see e.g. \cite{18})\footnote{We pay attention here only
on aspects associated with $\star$-structure of the theory}. As a
result one obtains
\begin{equation}\label{1}
\Gamma^{(1)}_\chi =\frac{1}{8\pi^2}\int d^8z \int^\infty_0 ds
s\,{\rm e}^{-s m^2} W_{\tilde{\star}}^2  \bar{W}_{\tilde{\star}}^2
\end{equation}
$$
\times \frac{\cosh(s{\cal N}_{\tilde{\star}})-1}{(s{\cal
N}_{\tilde{\star}})^2}\cdot \frac{\cosh(s\bar{\cal
N}_{\tilde{\star}})-1}{(s\bar{\cal N}_{\tilde{\star}})^2}
\cdot\frac{s^2({\cal N}_{\tilde{\star}}^2-\bar{\cal
N}_{\tilde{\star}}^2)}{\cosh(s{\cal
N}_{\tilde{\star}})-\cosh(s\bar{\cal N}_{\tilde{\star}})}~.
$$
The finite part of the
chiral contributions in the effective action can be written in
terms of the function $\zeta_{\tilde{\star}}$ (\ref{zetastar}):
\begin{equation}\label{2}
\Gamma^{(1)}_{ghosts} =\frac{1}{(4\pi)^2}\int d^8z \int^\infty_0 d
s \,{\rm e}^{-s m^2} W_{\tilde{\star}}^2 \bar{W}_{\tilde{\star}}^2
\zeta_{\tilde{\star}} (s{\cal N}_{\tilde{\star}}, s\bar{\cal
N}_{\tilde{\star}})
\end{equation}
Here $m$ is an infrared regulator mass.

As a result, the one-loop effective action on the covariantly
constant Abelian background is exactly calculated on the base of
manifestly gauge invariant techniques in the NAC superspace. We
emphasize a role of the star-operator $T^{\star}_{c}$ for the
theories with fields in the adjoint representation.

\section{Summary}
We have constructed a general procedure of calculating the
effective action for SYM theory coupled to matter in ${\cal
N}=1/2$ nonanticommutative superspace. The model is formulated in
terms of $\star$-product (\ref{st}) associated with the parameter
of nonanticommutativity ${\cal C}^{\alpha\beta}$ (\ref{hat}) and
preserves a half of initial ${\cal N}=1$ supersymmetry. The
effective action is formulated in framework of superspace
background field method.

We developed a proper-time techniques in ${\cal N}=1$ superspace
consistent with gauge invariance and $\star$-structure of the
theory under consideration. Superfield heat kernel determining the
structure of the one-loop effective action has been introduced for
the matter in the fundamental and adjoint representations for $SU(2)$
SYM theory.

The procedure for one-loop effective action calculation has been described.
We have applied this procedure to finding the low-energy effective
action for the matter in the fundamental representation in an external
constant Abelian vector multiplet background (\ref{onshell}) and for
the ${\cal N}=1/2$ SYM model with gauge group $SU(2)$
spontaneously broken down to $U(1)$. It was shown that in case of
matter in fundamental representation the low-energy effective action
(\ref{g1matter}) is obtained from the corresponding effective action
for the conventional ${\cal N}=1$ superfield theory by inserting the
$\star$-products instead of ordinary point-products. In case of SYM
theory, the effective action (\ref{1}), (\ref{2}) is also
constructed on the base of the effective action for the conventional SYM
theory, where however the products are given in terms of the special
star-operator $T^{\star}_{c}$ (\ref{t}) introduced in the paper. We
found that the models under consideration the effective action is
gauge invariant and written completely in terms of $\star$-product
and hence the classical $\star$-product does not get any quantum
corrections.

\section{Acknowledgements}
N.G.P is grateful to Sergei Kuzenko for his helpful comments and
suggestions. I.L.B is grateful to Trinity College, Cambridge for
finance support. Also he is grateful to DAMTP, University of
Cambridge and H. Osborn for kind hospitality. The work was
supported in part by RFBR grant, project No 03-02-16193. The work
of I.L.B was also partially supported by INTAS grant,
INTAS-03-51-6346, joint RFBR-DFG grant, project No 02-02-04002,
DFG grant, project No 436 RUS 113/669, grant for LRSS, project No
1252.2003.2. The work of N.G.P was supported in part by RFBR
grant, project No 05-02-16211.


\begin{thebibliography}{000}
\bibitem{kl}  D. Klemm, S. Penati, L. Tamassia,
Class.Quant.Grav. {\bf 20} (2003) 2905, hep-th/0104190; J. de
Boer, P. A. Grassi, P. van Nieuwenhuizen, Phys.Lett. {\bf B574}
(2003) 98, hep-th/0302078;  S. Ferrara, M. A. Lledo, O. Macia,
 JHEP {\bf 0309} (2003) 068, hep-th/0307039;
 H. Ooguri and C. Vafa, Adv.Theor.Math.Phys.
 {\bf 7} (2003) 53, hep-th/0302109;
 N. Berkovits and N. Seiberg, JHEP {\bf 0307}
(2003) 010, hep-th/0306226.


\bibitem{S}N. Seiberg, JHEP {\bf 0306} (2003) 010,
hep-th/0305248.

\bibitem{Iv}E. Ivanov, O. Lechtenfeld and B. Zupnik, JHEP {\bf 0402} (2004) 012,
hep-th/0308012; S. Ferrara and E. Sokatchev, Phys.Lett. {\bf
B579} (2004) 226, hep-th/0308021;
T. Araki, K. Ito and A. Ohtsuka, Phis. Lett. {\bf B606} (2005)
202, hep-th/0410203.

\bibitem{7}S. Terashima, J.-T. Yee, JHEP
{\bf 0312} (2003) 053, hep-th/0306237; M.T. Grisaru, S. Penati, A.
Romagnoni, JHEP {\bf 0308} (2003) 003, hep-th/0307099; R. Britto,
Bo Feng, S-J. Rey,  JHEP {\bf 0307} (2003) 067, hep-th/0306215;
JHEP {\bf 0308} (2003) 001, hep-th/0307091; R. Britto, B. Feng,
Phys.Rev.Lett. {\bf 91} (2003) 201601, hep-th/0307165; A.
Romagnoni, JHEP {\bf 0310} (2003) 016, hep-th/0307209; A.T. Banin,
I.L. Buchbinder, N.G. Pletnev, JHEP {\bf 0407} (2004) 011,
hep-th/0405063.



\bibitem{8}O. Lunin,
S.-J. Rey, JHEP {\bf 0309} (2003) 045, hep-th/0307275; D.
Berenstein, S.-J. Rey, Phys.Rev. {\bf D68} (2003) 121701,
hep-th/0308049; M. Alishahiha, A. Ghodsi, N. Sadooghi, Nucl.Phys.
{\bf B691} (2004) 111, hep-th/0309037.

\bibitem{9}I.Jack, D.R. Jones and L.A. Worthy, Phys.Lett. {\bf B611}
(2005) 199, hep-th/0412009; Phys.Rev. {\bf D72} (2005) 065002, hep-th/0505248.

\bibitem{10}S. Penati and A. Romagnoni, JHEP {\bf 0502} (2005)
064, hep-th/0412041.

\bibitem{11} S.J. Gates, Jr., M.T. Grisaru, M. Ro\v{c}ek and W. Siegel,
Superspace, Benjamin Cummings, Reading, MA, 1983.

\bibitem{12}I.L. Buchbinder and S.M. Kuzenko, Ideas and Methods
of Supersymmetry and Supergravity or a Walk Through Superspace,
IOP Publ. Bristol and Philadelphia, 1998.

\bibitem{13} S.M. Kuzenko, I.N. McArthur, JHEP {\bf 0305} (2003),
hep-th/0302205; JHEP {\bf 0310} (2003), hep-th/0308136.


\bibitem{16}T.D. Gargett, I.N. McArthur, Nucl.Phys. {\bf B497} (1997)
525, hep-th/9705200; N.G. Pletnev, A.T. Banin, Phys.Rev. {\bf D60}
(1999) 105017, hep-th/9811031; A.T. Banin, I.L. Buchbinder, N.G.
Pletnev, Nucl.Phys. {\bf B598} (2001) 371, hep-th/0008167.

\bibitem{18}I.L. Buchbinder, S.M. Kuzenko and A.A. Tseytlin,
Phys.Rev. {\bf D62} (2000) 045001, hep-th/9911221.

\bibitem{alv}L. Alvarez-Gaume, M.A. Vazquez-Mozo, JHEP {\bf 0504} (2005)
007, hep-th/0503016.


\bibitem{19}D.V. Vassilevich, Lett.Math.Phys. {\bf 67} (2004) 185,
hep-th/0310144;  JHEP {\bf 0508} (2005) 085, hep-th/0507123.




\end{thebibliography}
\end{document}